\documentclass[aps,prb,10pt,raggedbottom,twocolumn]{revtex4}

\usepackage{graphicx}
\usepackage{amsfonts}
\usepackage{times}

\begin{document}

\newcommand{\bra}[1]{\langle #1|\,} \newcommand{\ket}[1]{|#1\rangle}
\newcommand{\braket}[2]{\langle #1|#2\rangle}

\def\rplus{{c^+}}

\def\rminus{{c^-}}

\def\chargedensity{\rho}

\def\density{c}

\def\E{{\bf E}}

\def\B{{\bf B}}

\def\H{{\bf H}}

\def\Q{{\bf Q}}

\def\q{{\bf q}}

\def\p{{\bf p}}

\def\r{{\bf r}}
 
\def\J{{\bf J}}

\def\D{{\bf D}}

\def\C{{\bf C}}

\def\O{{\bf O}}

\def\r{{\bf r}}

\def\kt{{\, k_BT \, }}

\def\Ang{{\bf \Theta}}

\def\Inertia{{I}_{\theta}}

\def\Angvec{{\bf {\bf p}_{\theta}}}

\def\Angspeed{{{\bf v}_{\theta}}}

\def\dV{{\; \rm d^3}\r}

\def\curl{{{\,\rm curl}\; }}

\def\grad{{{\,\rm grad }\; }}

\def\div{{{\, \rm div }\; }}

\def\p{{\bf p}}

\def\rhat{\hat {\bf r}}

\title{Dynamics of a Local Algorithm for Simulating Coulomb
  Interactions} \author{A. C. Maggs} \affiliation {Laboratoire de
  Physico-Chime Th\'eorique, ESPCI-CNRS, 10 rue Vauquelin, 75231 Paris
  Cedex 05, France.  } \date{\today}
\begin{abstract}
  Charged systems interacting via Coulomb forces can be efficiently
  simulated by introducing a local, diffusing degree of freedom for
  the electric field. This paper formulates the continuum
  electrodynamic equations corresponding to the algorithm and studies
  the spectrum of fluctuations when these equations are coupled to
  mobile charges.  I compare the calculations with simulations of a
  charged lattice gas, and study the dynamics of charge and density
  fluctuations. The algorithm can be understood as a realization of
  a mechanical model of the ether.
\end{abstract}
\maketitle
\section {\bf Introduction}

Molecular dynamic simulation of charged condensed matter systems is
slow and difficult \cite{schlick}. In standard methods, such as
optimized Ewald summation \cite{ewald,ewald32}, fast multiple methods
\cite{multipole,greengard} or the fast Fourier transform
\cite{darden}, extensive and time consuming bookkeeping is needed
because of the range of the Coulomb interaction.  This bookkeeping
often scales badly when implemented on modern multiprocessor machines
which are used in the simulation of the largest systems. Naive
Monte-Carlo methods are particularly inefficient since the motion of a
single particle in an $N$ particle simulation requires the
recalculation of the Coulomb interaction with all other particles,
leading to a complexity in $O(N)$ for an update in the position of a
single particle

Recently, \cite{acm} a {\sl local algorithm}\/ with complexity scaling
as $O(1)$ per update was introduced for the Monte-Carlo simulation of
charged particles. In this algorithm an auxiliary electric field $\E$
is coupled to the charge density. The dynamics of $\E$ are chosen so
that the equilibrium distribution is determined by the Coulomb
interaction.  Due to the locality of the algorithm the method is
trivial to implement on multiprocessor machines.  In this paper we
study the dynamics of the algorithm in order to understand the
relaxation processes and time scales involved in a typical simulation.
Simulations are performed on a model of a charged lattice gas to
demonstrate the diffusive propagation of charge and density
fluctuations.

The algorithm is based on implementing Gauss's law
\begin{equation}
  \div \E = \chargedensity/\epsilon_0
\label{gauss}
\end{equation}
in the equivalent integral form
\begin{equation}
\int \E .\, d {\bf S} =         q/\epsilon_0
\label{integconstraint}
\end{equation}
as an exact dynamic constraint on the Monte-Carlo algorithm.  Here
$\chargedensity$ is the charge density and $\epsilon_0$ the dielectric
constant and $q$ the charge enclosed by the surface of integration in
eq.~(\ref{integconstraint}).


\begin{figure}
  \includegraphics[scale=.5] {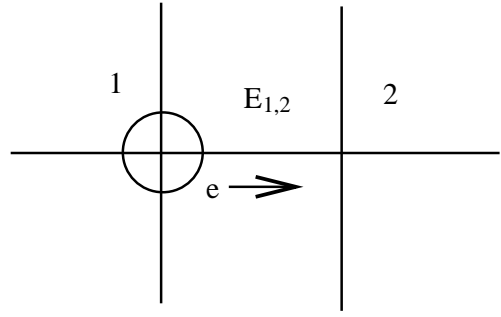}
\caption
{Motion of a charge $e$ from $1$ to $2$ is associated with a modification of
  the field on the connecting link: $E_{1,2} \rightarrow
  E_{1,2}-e/\epsilon_0$.}
\label{hydro1}
\end{figure}

\begin{figure}
  \includegraphics[scale=.5] {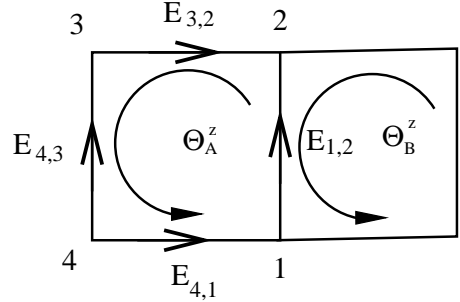}
\caption {Modification of the angle $\Theta^z_A$ leads to modified values of the
  field between on the links $\{1,2\}$, $\{3,2\}$, $\{4,3\}$ ,
  $\{4,1\}$.  The field in the $y$ direction associated with the link
  $E_{1,2}$ is given by the difference of the angles associated with
  the two plaquettes, $\Theta^z_A$ and $\Theta^z_B$, which are aligned
  in the $x$ direction. }
\label{hydro2}
\end{figure}

The system is discretized by placing charged particles on the vertices
of a cubic lattice, $\{i\}$. The electric field $E_{i,j}$ is
associated with the links $\{i,j\}$ of the lattice.  The simulation
starts with Gauss's law satisfied as an initial condition.  There are
two possible Monte-Carlo moves: Firstly, fig.~(\ref{hydro1}),
we displace a charge, $e$, situated on the leftmost lattice site, $1$,
to the rightmost site, $2$.  The discretized constraint is
\begin{equation}
\sum_{j} E_{i,j} = e_i / \epsilon_0 
\label{sum}
\end{equation}
with $e_i$ the charge at the site $i$.  The sum in eq.  (\ref{sum})
corresponds to the total electric flux leaving the site $i$. The
constraint is again satisfied if the field associated with the
connecting link is updated according to the rule $E_{1,2} \rightarrow
E_{1,2} - e/\epsilon_0$.  Secondly we update the field configurations,
fig.~(\ref{hydro2}),, by modifying the four field values of a single plaquette by a
pure rotation; $E_{1,2}$ and $E_{4,1}$ increase by an increment
$\Delta$ whereas $E_{4,3}$ and $E_{3,2}$ decrease by $\Delta$ so that
at each vertex the sum of the entering and leaving fields is
unchanged. The basic {\sl dynamic}\/ degree of freedom in the second
update is a circulation or rotation, $\Ang$, associated with each
plaquette of the network.

In the first section of the paper we formulate the continuum limit of
the evolution equations and show that they lead to diffusive evolution
of the electric field. We then couple the electric field to a mobile
gas of charged particles and compare the solutions of the coupled
plasma equations to simulations.  Finally we show that the equations
are closely related to the Maxwell electromagnetic theory.
\section{Diffusive Electrodynamics}
\subsection{Fundamental equations}

We start with a simple example to motivate our derivation of the
effective large scale equations obeyed by the electric field: a single
particle diffusing in a harmonic potential with energy ${\cal U} = K
x^2/2$.  The equation of motion is found by taking the derivative of
the energy with respect to the dynamic coordinate $x$ and then
balancing the resulting force against a relaxation process linear in
the velocity
\begin{equation}
\xi {dx \over dt} = - K x +f(t)
\label{simplelangevin}
\end{equation}
where $f(t)$ is an external forcing term. $\xi$ the inverse mobility
sets the characteristic time scale of the relaxation of $x$ and is a
function of the step size of the Monte-Carlo trial moves.  A first
order (in time) algorithm, such as Monte-Carlo, for the simulation of
a particle in such a potential (performed near equilibrium with small
step sizes) is essentially a discretized realization of this
stochastic differential equation.  The Langevin description is
completed by specifying $f$, so as to obey the fluctuation dissipation
theorem.

We now turn to the equations for the electric field. Firstly we
examine the field in absence of current then we shall find the
coupling of the field to external sources.  The discretized energy of
the electric field is given by
\begin{equation}
{\cal U}={\epsilon_0\over 2}\sum_{links} E_{i,j}^2 \quad .
\end{equation}
The basic variables in the dynamics of the field are, however, not
$\E$ but rather the rotational degrees of freedom which are updated
independently at each time step. The conjugate force acting on this
variable is a torque.  We define on each plaquette in the $\{x,y\}$
plane the angular variable $\Theta^{z}$. We group the angle
corresponding to the three possible plaquette orientations into a
vector ${\bf \Ang}$.  The physical field $\E$ is sensitive to
derivatives of ${\Ang}$: Studying fig. (\ref{hydro2}) one sees that
$E_{1,2}$ is given by the $x$ variation of the $z$ component of
$\Ang$. This we recognize as part of the curl operator acting on the
field $\Ang$.  The complete relation between a local variation in $\E$
and a local variation in $\Ang$ is thus
\begin{equation}
\delta \E =  \curl \delta \Ang
\label{first}
\end{equation}
where in the discretized model derivatives are to be understood as
finite differences with $\E$ living on links and $\Ang$ on plaquettes.

Modification of the circulation $\Theta^z_A$ in fig. (\ref{hydro2})
 by $\Delta$ gives rise to a new contribution to the energy of the
plaquette.
\begin{eqnarray}
\bar {\cal U}= 
{\epsilon_0 \over 2} \,  (&(&E_{4,1} + \Delta)^2 +
(E_{1,2} + \Delta)^2 +\\
&(&E_{4,3} - \Delta)^2 +
(E_{3,2} - \Delta)^2 )  \nonumber
\end{eqnarray}
Taking the derivative with respect to $\Delta$ we find a torque, $\C$,
acting on the circulatory degree of freedom which is a discretized
version of the curl of the field.
\begin{eqnarray}
C_z = -{\partial \bar {\cal U} \over \partial \Delta}&=& -\epsilon_0 \left( \, (E_{4,1} -E_{3,2}) + 
(E_{1,2} -E_{4,3}) \right)\\
&=& -\epsilon_0 \left( {\partial E^y \over  \partial  x } 
- {\partial E^x \over  \partial  y } \right )\\
&=& -\epsilon_0 \hat {\bf k}\cdot  \curl \E 
\label{second}
\end{eqnarray}
Again the three components of the torque live on the plaquettes
together with the angle $\Ang$.  By analogy with eq.
(\ref{simplelangevin}) the evolution equation for the angle $\Ang$ is
\begin{equation}
\xi {d \Ang \over dt} =  \C = -\epsilon_0 \curl \E
\label{third}
\end{equation}
linking a velocity to the conjugate force.  Eventually a stochastic
force should also be added into this equation but we shall not need it
in what follows.

Consider now the evolution of the field in the presence of a current.
Take a network in which charges are present at {\sl every}\/ vertex
and displace every charge to the right as in fig.~(\ref{hydro1}).
  Then every bond in the $x$ direction is modified by
$-e/\epsilon_0$ where $e$ is the charge displaced even though the
local charge density is unchanged. If we displace the charges at a
constant rate we have the evolution of the field due to the source as
\begin{equation}
{\partial \E_{\rm source} \over \partial  t} = -\J/\epsilon_0
\label{fourth}
\end{equation}
Combining eqs. (\ref{first},\ref{fourth}) we find
\begin{equation}
{\partial \E_{\rm} \over \partial  t} 
= -\J/\epsilon_0 + \curl { {\partial \Ang \over \partial t }}
\label{six}
\end{equation}
This equation is clearly analogous to the Maxwell equation,
\begin{equation}
\epsilon_0 {\partial \E_{\rm} \over \partial  t} 
= -\J + \curl \H
\end{equation}
if we write that $\H= \epsilon_0 { {\partial \Ang \over \partial t
    }}$. We shall see later that $\H$ is indeed closely related to the
magnetic degrees of freedom of classical electrodynamics. From eqs.
(\ref{second},\ref{third},\ref{six})

\begin{equation}
  {\partial \E \over \partial t} = 
 \left (\epsilon_0 \nabla^2 \E - \grad \chargedensity \right)/\xi  -\J/\epsilon_0 
\label{basic}
\end{equation}
where we have used the standard identity $\curl \curl \E=( \grad \div
\E- \nabla^2 \E)$ and Gauss's law. Equation (\ref{basic}) is the main
result of this section giving a diffusive propagation law of the
electric field in absence of external charges and currents.

In the static limit both the current and the time derivative of eq.
(\ref{basic}) vanish.  We find the same equation for the electric
field
\begin{equation}
 \nabla^2 \E = \grad {\chargedensity \over \epsilon_0}  
\end{equation}
as is found by applying the operator $(-\grad)$ to the Poisson
equation in conventional electrostatics. When we take the divergence
of eq. (\ref{basic}) we discover that
\begin{equation}
\left ( 
{\partial \over \partial t} - {\epsilon_0\over \xi} \nabla^2 \right ) (\div \E -\chargedensity / \epsilon_0)=0 \label{diffgauss}
\end{equation}
Again we see that Gauss's law is implemented in the method as an
initial condition. Note that for this to be true we require that the
Langevin noise associated with  eq. (\ref{basic}) does not
in itself destroy the conservation law. It is thus the $\curl$ of some
vector field \cite{chaikin}.

\subsection{Relation to Potentials}

In our earlier paper \cite{acm} we showed that the electric field
could be calculated from a scalar potential $\phi$ and a vector
potential $\Q$ with the relationship
\begin{equation}
\E = -\grad  \phi + \curl \Q
\label{potentials}
\end{equation}
In Fourier space we can write this equation as
\begin{equation}
  \E({\bf k}) = -i {\bf k} \phi + i {\bf k} \wedge \Q
\end{equation}
The second term of this expression is perpendicular to ${\bf k}$ so
that there are two transverse degrees of freedom in the $\Q$ field,
corresponding to two independent polarization states; the longitudinal
component of $\Q$ is projected out and does not contribute to the 
electric field.  We can consider
that the field is due to a static longitudinal potential plus
transverse photons.

If we take the time derivative of eq. (\ref{potentials}) we can
compare with eq. (\ref{six}). The term $\curl \dot \Ang$ is purely
transverse, whereas $-\J$ contains both longitudinal and transverse
components. Thus we conclude that
\begin{equation}
\curl \dot \Q = \curl \dot \Ang - \J_{t}/\epsilon_0
\end{equation}
and
\begin{equation}
\grad \dot \phi = \J_l/\epsilon_0
\end{equation}
where $\J_t$ and $\J_l$ are the transverse and longitudinal components
of the current. In general these are {\sl non-local} relationships
since the projection  involves a passage via Fourier
components. Such non-local relationships between potential and field
are normal in the Coulomb gauge \cite{jackson}.

\subsection{Phenomenological Dynamics of a Two Component Plasma}

In this section we couple the diffusive evolution equation for the
electric field eq. (\ref{basic}) to the equations describing a two component plasma
and study the relaxation phenomena and time scales that are to be
expected when using the algorithm to simulate dense charged systems.

The equations of conservation and linear response give the following
equations for the charge degrees of freedom:
\begin{eqnarray}
{\partial \rplus \over \partial t} &=& -\div {\J^+} \nonumber \\
{\partial \rminus \over \partial t} &=& -\div {\J^-} \nonumber \\
\J^+ &=& -D \grad \rplus +e \mu \rplus \E \nonumber \\
\J^- &=& -D \grad \rminus - e \mu \rminus \E  \label{cons}
\end{eqnarray}
Where $\rplus$ and $\rminus$ are the number densities of of the
positive and negative charges, $\pm e$.  $\J^\pm$ are the number
current density.  From these equations we find the equations obeyed by
the total density $\density$ and the charge density $\chargedensity$.
We note that the diffusion coefficient $D$ and the mobility, $\mu$ are
linked by the Einstein relation $D=\kt \mu$.

Taking the sum and difference of the equations (\ref{cons}) we find
\begin{eqnarray}
{\partial \density  \over \partial t} &=& D \nabla^2 \density \label{diffusion} \nonumber \\
{\partial \chargedensity  \over \partial t} &=& -\div \J \nonumber \\
\J &=& e^2\mu \density_0 \E - D \grad \chargedensity - e^2 c_0 \mu \grad \phi_e(t) \label{current}
\label{charge}
\end{eqnarray}
where we have linearized the equations about a mean density
$\density_0$. $\J= e \J^+ - e \J^-$ is the electric current density.
$\phi_e(t)$ is an externally imposed potential that we shall use to
calculate charge-charge correlation functions.  Substituting eq.
(\ref{current}) for $\J$ in (\ref{basic}) we find
\begin{eqnarray}
\label{plasma}
  {\partial \E \over \partial t} &=& 
 \epsilon_0/\xi\, \nabla^2 \E - 
 e^2 \mu \density_0/\epsilon_0\, \E  \\ 
 &+& (D/\epsilon_0 -1/\xi)\, \grad \chargedensity + e^2 c_0 \mu/\epsilon_0\, \grad \phi_e \nonumber
\end{eqnarray}
We analyze this equation by treating separately the longitudinal and
transverse fluctuations. Take the $\curl$ of eq.  (\ref{plasma}) to
find that the transverse components of $\E$ decouple from the charge
density. In Fourier space we find the dispersion law
\begin{equation}
(i  \omega + \epsilon_0/\xi\, q^2 +e^2 \mu \density_0/\epsilon_0)\; 
{\bf q \wedge \E}=0
\end{equation}
In the absence of charges the mode is diffusive but the presence of a
finite charge density gives a gap in the spectrum.

Consider now the equations for the field eq. (\ref{plasma}): With the
help of Gauss's law one replaces the divergence of the field by the
charge to find
\begin{equation}
\left( {\partial \over \partial t} - D \nabla^2 
+ e^2 \mu \density_0/\epsilon_0 \right ) \chargedensity = c_0 e^2\mu \nabla^2 \phi_e
\label{PB}
\end{equation}
which also applies to the longitudinal mode of the electric field
\begin{equation}
( i \omega + D q^2 +e^2 \mu \density_0/\epsilon_0)\;
 {\bf q}\cdot \E = -e^2 c_0 \mu q^2 \phi_e/\epsilon_0
\end{equation}
Again the spectrum has a gap as $q\rightarrow 0$.

\section{Numerical Results}
\subsection{Dynamics}

We performed simulations of a charged lattice gas to study the
dynamics of the density and charge fluctuations. Equal numbers of
positively and negatively charged particles with $e=\pm 1$ were placed
on the vertices of a network which was simulated by the algorithm in a
uniform dielectric background.  During the simulations we measured the
Fourier transform of the particle distributions
\begin{equation}
s({\bf q} ,t)=   {1\over \sqrt{N}}
\sum_i e_i \exp(i {\bf r}_i (t)\cdot {\bf q}_i )
\end{equation}
where the weight $e_i$ is the charge for the charge correlation
function and is unity for the density correlation function.  We use
this information to construct the dynamic structure factor
\begin{equation}
  S( {\bf q},t) = \langle s({\bf q} ,t)  s({\bf -q} ,0) \rangle
\label{sqt}
\end{equation}
The result is fitted with an exponential and the decay rate plotted as
a function of $q^2$ in figs.
(3,4).  The density-density correlation function displays simple
diffusive behavior.  The charge-charge correlation function is
characterized by a gap at $q=0$.

\begin{figure}
  \includegraphics[scale=.47] {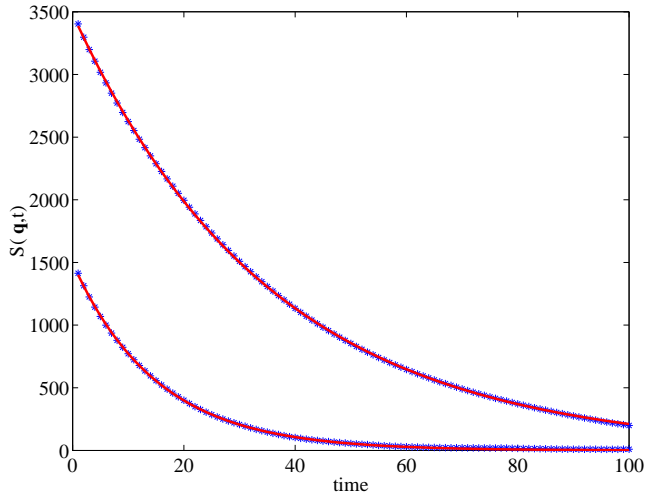}
\caption {Fit of a density-density, top, and a charge-charge, bottom,
  correlation functions eq. (\ref{sqt}) to a single exponential. 5000
  particles, network of $25\times25\times25$, mode $\q= 2\pi\times
  (2,2,0)$.  Arbitrary units}
\label{fit}
\end{figure}

What do these dispersion relations imply for the equilibration of a
system of charged particles?
The mass degree of freedom is diffusive so that a simulation
equilibrates in a time which scales quadratically with the linear
dimensions of the system. The charge degree of freedom is associated
with a Green function which is also diffusive. However, the total
weight decays exponentially in time. The signal due to a charge
fluctuation is very weak beyond the Debye length.

\begin{figure}
  \includegraphics[scale=.47] {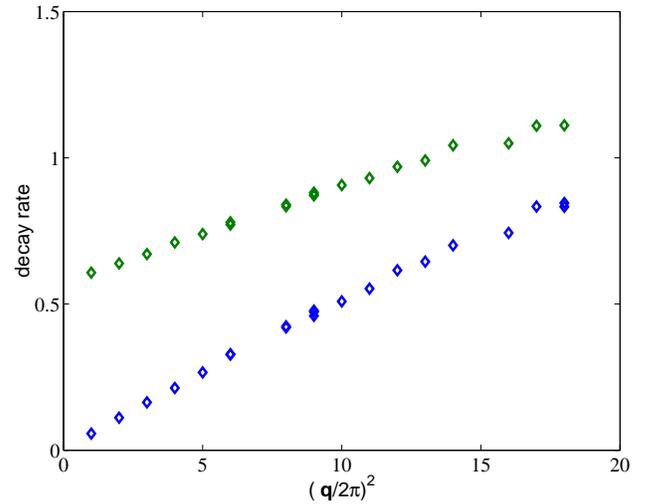}
\caption {Characteristic time extracted from $S(q,t)$ as a function of $q^2$.
  Bottom curve : Density-density correlations: the mode is
  diffusive, eq.(\ref{current}).  Top curve: The charge-charge
  correlation function has a gap in agreement with eq. (\ref{PB}).
  Selected modes between $(q/2\pi)^2=1$, $\q= 2\pi\times (1,0,0)$ and
  $(q/2\pi)^2=18$, $\q=2\pi\times(4,1,1)$.  1000 Particles, $18\times
  18\times 18$ mesh. Vertical axis in arbitrary units. }
\label{times}
\end{figure}

Note that the parameters used in the derivation of the plasma dynamics
are already at a coarse grained level of description. We expect that
the bare parameters are renormalized by non-linear interactions: While
the eq.  (\ref{basic}) is in some sense fundamental, containing within
it the exact statement of Coulomb's law and Boltzmann statistics, the
eqns (\ref{current}) are purely phenomenological.  An example is the
mobility of a particle $\mu$ which in the above theoretical
presentation appears independent of the field parameters $\epsilon_0$
and $\xi$. However consider the case of a charged particle pulled by
an external non-electric force in the presence of a electric field
which relaxes very slowly. As the particle moves it leaves behind it a
``string'' of electric field due to the dynamics of fig.~(\ref{hydro1}).
  This creates a back force on the particle which reduces its
mobility.  Monte-Carlo moves on the field spread this string over many
lattice sites increasing the mobility of a charged particle. Thus the
mobility of the charged particles increases when the field relaxes
more rapidly.

This effect is an explanation of the curves of fig.~(\ref{times})
 Despite the predictions of eqs. (\ref{charge},\ref{PB}) the slope
of the charge-charge and the density-density curves are slightly
different; the effective diffusion coefficient of the charge
fluctuations is lower than that of the density fluctuations.  Slow
relaxation of the electric degrees of freedom should hinder the motion
of a single charged particle more than a strongly coupled, neutral
pair moving in the same direction.

\subsection{Screening}

\begin{figure}
\includegraphics[scale=.47] {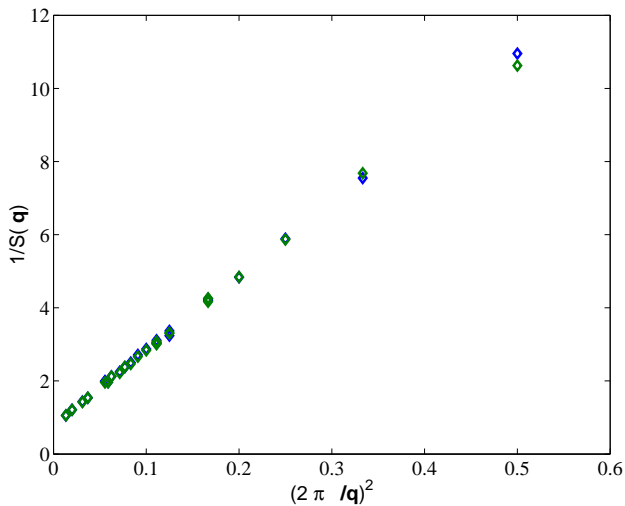} 
\caption {Plot of $1/S(q)$ as a function of $(2\pi/q)^2$. The plot is linear
  as implied by eq. (\ref{lorentz}). Selected modes between $2\pi\times(1,1,0)$
  out to $2\pi\times(5,5,5)$.  The plasma screens interactions exponentially.
  5000 charges on a network of $25\times25\times25$ }
\label{sq}
\end{figure}

From the Poisson-Boltzmann equation it is known that charged systems
screen. We derive this result from our dynamic equations as follows:
Consider eq.  (\ref{PB}) for the charge density in the presence of a
static external potential $\phi_e(\q)$.
\begin{equation}
\chargedensity ( \q ) = {-q^2 \over q^2 + e^2 \density_0/\epsilon_0\kt} {c_0 e^2\phi_e \over \kt}
\end{equation}
from linear response theory the structure factor with the
normalization of eq. (\ref{sqt}) is given by
\begin{equation}
S(q) = {e^2 q^2 \over \kappa^2 +q^2}
\label{lorentz}
\end{equation}
where the inverse Debye length, $\kappa$, is given by the standard expression
$\kappa^2 = e^2 \density_0/\epsilon_0 \kt$.  This prediction is
checked in our code by plotting $1/S(q)$ as a function of $1/q^2$,
fig.~(\ref{sq}).

Fig.~(\ref{sq}) 
 should be taken as very strong evidence that our algorithm is
behaving correctly. It reproduces one of the most striking features of
charged systems, the exponential decay of the charge-charge
correlation function due to Debye screening.

\subsection{Numerical Stability}
In the simulations that we performed to study the dynamic and
screening properties of the algorithm we were agreeably surprised by
the numerical stability of the algorithm: At each update one makes an
error $e_p$ comparable to the round off error of the computer.  Over
many time steps this accumulates so that Gauss's law is violated.  We
feared that this local error would rapidly become important.
 
The slow propagation (in time) of numerical errors can be understood
by consideration of eq.  (\ref{diffgauss}). Local fluctuations in the
constraint $\div \E -\chargedensity/\epsilon_0=0$ spread out via a
diffusion process. Since both positive and negative errors are made
during a simulation there is a large degree of cancellation occurring.
After a single Monte-Carlo sweep of the system Gauss's law is violated
by $O(e_p)$ at each lattice site.  However averaged over a sample with
$L^3$ sites the average error per site is $O(e_p/L^{3/2})$.  When
simulated for $O(L^2)$ sweeps the system comes to equilibrium under
the diffusive propagation of the charge and density fluctuations, we
find errors of only $O(e_p/ L^{1/2})$ per site.  The high statistics
curves of this paper were generated by using runs of length up to $5000$
times the equilibration time.  Even here the errors remained
acceptably small.

\section{Propagative Field Equations}
\subsection{Maxwell's equations}
In eq. (\ref{basic}) we gave the equations of motion for the electric
field obeyed in the continuum limit of a Monte-Carlo simulation. In
this section we shall see how local imposition of Gauss's law can be
used to find a propagative dynamics for the evolution of the electric
field. We continue to describe the basic dynamic degree of freedom as
an angular variable, $\Ang$, which is linked to the electric field by
eq. (\ref{six}). This variable is associated with a angular velocity,
$\Angspeed$ and moment of inertia $\Inertia$. The resulting second
order equations will display propagating, wave like features rather
than the diffusive propagation characteristic of eq. (\ref{basic}).

Each link field $E_{i,j}$ is the result of the rotation of a variable
$\Ang$ defined on the faces of the cube rotating at angular velocity
$\Angspeed$.  As above the torque on the rotational degree of freedom
of a plaquette is given by
\begin{math}
  \C = -\epsilon_0 \curl \E
\end{math}.
Using eq. (\ref{first},\ref{fourth}) we find the following equations
obeyed by the fields:
\begin{eqnarray} 
\Inertia {\partial \Angspeed \over \partial t}&=& -\epsilon_0 \curl \E \nonumber \\
{\partial  \E \over \partial t} &=& \curl \Angspeed -\J/\epsilon_0 \nonumber \\
\div \E &=& \chargedensity/\epsilon_0 \label{quasimaxwell}
\end{eqnarray}
where the differential operators are to be interpreted as the
appropriate difference when acting on the lattice variables.  The
equations (\ref{quasimaxwell}) are a rescaled version of Maxwell's
equations with $\Angspeed$ playing the role of the magnetic field
${\bf H}$. 

In order to find the coupling between  particles and the variables
$\Ang$ we are obliged to use the formalism of Lagrangian dynamics.
Naive arguments based on energy considerations are ambiguous and can
easily lead to wrong results.

\subsection{Lagrangian Treatment of Dynamics}
We shall now show how to derive the full coupled equations between
particle motion and field. Firstly, however, we shall look at a simple
illustrative example in order bring out the main formal features of
constrained Lagrangian dynamics.

Consider two gears described by the rotation angles $\varphi$ and
$\psi$. We take these gears to have unit inertia and impose on them a
potential energy $g(\varphi)$ and $h(\psi)$.  The gears are in contact
and are thus submitted to the rolling constraint
\begin{equation}
\dot \varphi + \dot
\psi =0
\end{equation}
We find that the Lagrangian describing this system is simply
\begin{equation}
{\cal L} = {\dot \varphi^2 \over 2} +  {\dot \psi^2 \over 2} -g -h +
A (\dot \varphi + \dot \psi)
\end{equation}
where the Lagrange multiplier $A$ imposes the constraint. Note that we
are {\sl not} using the standard method of D'Alembert of imposing
nonholonomic constraints but rather the ``vakonomic'' method
\cite{arnold} in which the field $A$ is itself considered an
independent dynamic variable. Such methods are widely used in field
theory, see for instance the book of Schwinger \cite{schwinger}.

From this Lagrangian we find the equations of motion and the momenta.
For instance
\begin{equation}
p_\varphi =  \dot \varphi  +A
\end{equation}
and
\begin{equation}
{d^2 \varphi \over d t^2} = - {d g \over d \varphi} - {d A \over d t}
\end{equation}
These equations linking the momentum $p_\varphi$ to the velocity and
the equation for the acceleration of the variable $\varphi$ are
remarkably similar to those found in electromagnetism if one
interprets $A$ as the vector potential.

We shall now use the same trick of considering the constrained
Lagrangian dynamics of the field $\Ang$ to find the coupled equations
for the field and moving particles.  We interpret the variable $\dot
\Ang$ as a rotation velocity and  $\E^2$ as a
potential energy. For notational simplicity we consider unit mass
particles and a system of units where $\epsilon_0=I_\Ang=1$.  We find
the following Lagrangian:
\begin{eqnarray}
{\cal L} &=& 
\sum_i {\dot \r_i^2 \over 2}+
\int d^3\r
\left( {{\dot \Ang}^2 \over 2} -{\E^2 \over 2} \right) \\ 
&+& \int d^3\r \; {\bf A}\cdot ( 
 \dot \E - \curl \dot \Ang + 
\sum_i q_i \delta(\r - \r_i)  \dot \r_i )\nonumber
\end{eqnarray}
Here the Lagrange multiplier $\bf A$ imposes the kinematic constraint
eq. (\ref{six}) in a manner analogous to the rolling constraint for
the gears.  We find the equations of motion by the usual variational
calculus: it is useful to note that the $\curl$ operator is self
adjoint with appropriate boundary conditions so that
\begin{math}
  \int {\bf A} \cdot \curl {\bf B} \dV= \int {\bf B} \cdot \curl {\bf
    A} \dV
\end{math}
\begin{eqnarray}
\label{variations}
\delta \Ang &:&\quad  \ddot \Ang - \curl \dot {\bf A} =0\\ \nonumber
\delta \E   &:&\quad \E + \dot {\bf A}=0\\ \nonumber
\delta \r_i &:&\quad \ddot \r_i + q_i {d {\bf A}\over dt} - q_i \grad(\dot \r_i.{\bf A}) =0\\ \nonumber
\delta {\bf A} &:& \quad \dot \E - \curl \dot \Ang + 
\sum_i q_i \delta(\r - \r_i)  \dot  \r_i  =0
\end{eqnarray}

The variation in $\delta \r_i$ can be rewritten by using the identity
\begin{math}
  \grad ({\bf v.A}) = ({\bf v}.\grad) {\bf A} + {\bf v} \wedge
  \curl{\bf A}
\end{math}
and by noting that
\begin{math}
  {d \over dt} = ({\partial \over \partial t} + {\bf v}. \grad)
\end{math}
\begin{eqnarray}
\ddot \r_i + q_i {\dot {\bf A}} -q_i \dot \r_i \wedge \curl {\bf A} &=0& \nonumber\\
\ddot{ \r}_i =  q_i(\E + \dot \r_i \wedge \curl \dot \Ang)
\end{eqnarray}
These are the normal equations of electromagnetism if we identify
$\dot \Ang$ with $\B$. The Lagrangian corresponds to the {\sl temporal
  gauge} where the scalar potential $\phi=0$. A gauge transformation
${\bf A} \rightarrow {\bf A} + \grad \Psi(t)$ generates additional terms in the Lagrangian
of the form $\phi( \div \E - \rho)$ with $\phi= \dot \Psi$.

We can eliminate $\dot \Ang$ from the Lagrangian via the Thomson-Routh
treatment of kinesthenic variables: Consider the modified action
\begin{math}
  \bar {\cal L} = {\cal L} - \p_{\theta} \dot \Ang
\end{math}.
We find that
\begin{equation}
\bar {\cal L} = 
\sum_i {\dot \r_i^2\over 2}-
\int d^3\r
\left( {({\curl \bf A})^2\over 2} +{\E^2 \over 2} \right)  
+ \int d^3\r \; {\bf A}\cdot (\dot \E  +  \J ) \nonumber
\end{equation}
which we recognize \cite{schwinger} as a more conventional Lagrangian
for electrodynamic systems.

We construct the Hamiltonian using Dirac's procedure with two
constraints:
\begin{eqnarray}{\cal H} &=&
\sum_i   {( \p_i -q_i {\bf A}(\r_i))^2 \over 2} 
+ \int   {( \p_\theta + \curl {\bf A})^2 \over 2}\; d^3\r  \nonumber \\ 
  &+& \int \left( {\E^2 \over 2}
+ \mu {\bf p}_{\bf A} + \gamma ({{\bf p}_\E -{\bf A}}) \right)\; d^3 \r
\end{eqnarray}
where $\mu$, $\gamma$ are the multipliers for the primary constraints.
The initial conditions are ${\bf p}_E={\bf A}$, ${\bf p}_A=0$ and
${\bf p_\theta}=0$ which are conserved by the equations of motion in
the same way that Gauss's law is conserved in the Monte-Carlo
formulation. On the physically relevant surface the constraint terms
are identically zero; the extended Hamiltonian still has the normal
interpretation as the conserved total energy.

Such a description of the electromagnetic field in terms of rotors was
known to FitzGerald in the nineteenth century as a mechanical analogy
\cite{model,maxwellians} of the ether. A square array of wheels was
constructed; neighboring wheels were connected by an elastic band.
When two neighboring wheels turn at the same angular velocity the
elastic band has constant length and the elastic energy is constant.
When there is a difference of rotational velocity between wheels the
elastic energy of the bands changes. Assuming linear elasticity for
the elastic bands one finds an exact mapping of electromagnetism onto
a mechanical problem. This model was most important in the history of
electromagnetism: FitzGerald used this model in the very first
calculation of radiated power from moving charges.

\subsection{Statistical Mechanics}

The interpretation of $\dot \Ang$ as the angular velocity of a rotor
suggests that it could be coupled to a thermostat to improve
equilibration of the field degrees of freedom; the linear equations
that we have found for the electric field are likely to equilibrate
rather slowly. If we add coupling to external noise, $\vec \zeta$ and
friction, $\Gamma$, in eq. (\ref{variations}) we find
\begin{equation}
\ddot \Ang = \curl \dot {\bf A} - \Gamma \dot \Ang + \vec \zeta(t) \label{damp}
\end{equation}
thus coupling the angular velocity to an arbitrary thermostat leads to
violation of the Maxwell equation $\div \B=0$.

The partition function is calculated from
\begin{equation}
{\cal Z} = \int {\cal D}{\bf p} \; 
  {\cal D}{\bf q}\;  e^{-\beta {\cal H}}
\end{equation}
The integral is over the canonical coordinates ${\bf q}$ and momenta ${\bf p}$.
The integration region is the set of configurations available to the
equations of motion.  We thus implicity include delta function constraints on $\p_E$ and $\p_A$.
Integration over the momenta is easy to perform
in the presence of Langevin noise  which destroys the
constraints and conservation laws associated with the variable
$\Angvec$ in Maxwell's equations. What remains is the integral over
the electric fields and particle positions. If the dynamics were
ergodic we would integrate over all values of the field.  However
Maxwell's equations, even in the presence of noise on the momentum
degree of freedom, include Gauss's law. This constrains the electric
field and the partition function is given by
\begin{equation}
{\cal Z}_c = \int {\cal D}\r_i\,
\int {\cal D} \E\, e^{-\int {\beta \epsilon_0 \E^2\over 2 } \dV}\,
 \prod_{\bf r} \delta(\div \E - \rho/\epsilon_0) \label{final}
\end{equation}
where, now, all degrees of freedom are freely integrated over.  It is
this constrained configurational integral \cite{acm} that leads to
effective Coulomb interactions.

Combining eqs. (\ref{six}) and (\ref{damp}) we find the equation for
the electric field:
\begin{equation}  {\partial \over \partial t}
\left( {\partial \over \partial t} +\Gamma \right) \E= \nabla^2 \E - \grad \rho -
\left( \Gamma +  {\partial \over \partial t} \right) \J + \curl \vec \zeta(t) \label{largescale}
\end{equation}
In the limit of low frequencies we can ignore ${\partial \over
  \partial t}$ compared to $\Gamma$ and find an equation entirely
equivalent to eq. (\ref{basic}). The damping strongly modifies the
large scale nature of the electric field dynamics.

This result seems quite remarkable. It is known from the work of
Heaviside \cite{heaviside} that the electric field of a moving
particle is strongly modified at velocities which approach $c$, the
speed of light. Despite this eq.  (\ref{final}) implies that the
average interaction between particles is independent of this
longitudinal contraction of the electric field.  Either this is
consequence of the full Maxwell equations which has not yet been or
explored or it is a consequence of relaxing the constraint on the
divergence of the magnetic field leading to the modified large scale
properties implied in eq. (\ref{largescale}). We leave further study
of this problem to a future publication. In either case this could
permit study of the Coulomb interacting particles via direct integration of the
Maxwell equations in molecular dynamics simulations.

\section{Conclusions}
We have analyzed a Monte-Carlo algorithm for the simulation of long
ranged Coulomb interactions. We have seen that propagation of the
electric degrees of freedom is diffusive. By construction the
dynamics sample the equilibrium Boltzmann distribution of the charged
system.  The locality of the algorithm allows fast and {\sl simple}\/
implementations even on multiprocessor computers with high
communication overheads. We have verified that the Monte-Carlo
algorithm reproduces well known features of the two component plasma
such as screening.
Our law for the {\sl local}\/ update of the electric field
after movement of a particle, fig
(1), is a discretized version of the Maxwell displacement current.
The algorithm has been shown to be closely related
to mechanical models of the ether introduced in the generation that
followed Maxwell.

\bibliography{mc}

\end{document}